\documentclass{aa}

\usepackage{tabularray}
\usepackage{geometry}
\usepackage{multirow}
\usepackage{booktabs}
\usepackage{graphicx}
\usepackage{txfonts}
\usepackage{lipsum}
\usepackage{amsmath}
\usepackage{siunitx}
\usepackage{natbib}
\usepackage{hyperref}
\usepackage{subcaption}         
\usepackage{lscape}             
\usepackage{placeins}           
\usepackage{scalerel}
\usepackage{tikz}

\begin{document}
\makeatletter
\let\linenumbers\relax
\let\runninglinenumbers\relax
\let\pagewiselinenumbers\relax
\let\columnwiselinenumbers\relax
\makeatother

   \title{Multibanded Reduced Order Quadrature Techniques for Gravitational Wave Inference}


   \author{
Murdoc Newell\inst{1}
\and
Alexis Boudon\inst{2}
\and
Hong Qi\inst{1}\thanks{Corresponding author: hong.qi@ligo.org}
}

   \titlerunning{Accelerating ROQ construction with multibanding}
   \authorrunning{Newell et al.}

   \institute{School of Mathematical Sciences, Queen Mary University of London, 327 Mile End Road, London, E14NS, United Kingdom  \and Univ Lyon, Univ Claude Bernard Lyon 1, CNRS/IN2P3, IP2I Lyon, UMR 5822, F-69622, Villeurbanne, France\\ }


 
\abstract{Reduced-order quadrature (ROQ) is commonly used to accelerate parameter estimation in gravitational wave astronomy; however, constructing ROQ bases can be computationally costly, particularly for longer-duration signals. We propose a modified construction strategy based on PyROQ that accelerates this process by performing the basis search using multiband waveforms, without compromising the desired likelihood accuracy. We use this altered method to construct a set of ROQs in the sub-solar mass (SSM) range using the \texttt{IMRPhenomXAS\_NRTidalV3} waveform. Compared to PyROQ's standard ROQ method, we find a decrease in basis size of 20\% to 30\% and observe a decrease in basis construction time by 5 to 20 times, reducing from two weeks to a couple of days.  We verify the bases built with this method by injecting simulated gravitational waves into LIGO-Virgo-KAGRA design noise and recovering the parameters, and we find that they preserve the likelihood accuracy and maintain consistent parameter estimation results.}
\keywords{gravitational waves -- reduced order quadratures --  multibanding}
\maketitle

\section{Introduction}
The direct detection of gravitational waves has opened a new window into the Universe, enabling the study of compact binary coalescences and other extreme astrophysical phenomena. Since the first detection of a binary black hole merger in 2015 \citep{PhysRevLett.116.061102}, the LIGO-Virgo-KAGRA detector network has observed hundreds of gravitational wave events, with compact binary mergers being the dominant source class \citep{LIGOScientific:2018mvr, LIGOScientific:2020ibl, KAGRA:2021vkt, LIGOScientific:2025slb}. Each detection requires sophisticated parameter estimation techniques to infer the physical properties of the source systems.

Bayesian parameter estimation for gravitational wave signals has been traditionally used to evaluate the likelihood function millions of times across the multi-dimensional parameter space \citep{Veitch:2014wba, Pankow:2015cra, Lange:2018pyp, Ashton2019, Smith:2019ucc, Wouters:2024oxj, Dax:2024mcn}. For each likelihood evaluation, the overlap integral between the observed detector data and theoretical waveform templates must be computed across the effective frequency band of the detectors, typically from tens to thousands of Hertz. The computational cost scales linearly with the number of frequency samples and nonlinearly with the complexity of the waveform models. For longer-duration signals, such as those from subsolar-mass (SSM) compact objects that remain in the detector sensitivity band for extended periods, this computational burden becomes prohibitive, with individual parameter estimation analyses requiring weeks or even months of computing time.

Reduced order quadrature (ROQ) techniques have emerged as a powerful solution to this computational challenge \citep{Canizares:2014fya, Smith:2016qas, PyROQ, PhysRevD.108.123040, PhysRevD.108.123025}. By constructing reduced bases that accurately represent the space of possible waveforms, ROQ techniques enable the overlap integrals to be evaluated using only a small subset of frequency points, achieving speedups of two to three orders of magnitude while maintaining likelihood accuracy to within acceptable tolerances. Our previous code, PyROQ, implements an efficient algorithm for constructing these reduced bases.

However, the construction of ROQ bases itself presents a computational bottleneck. The search algorithm requires generating large training sets of waveforms, often millions of templates, and performing numerous inner product evaluations at each iteration. For long-duration signals like SSMs with hundreds of thousands of frequency samples, this construction process can take days to weeks on high-performance computing clusters, limiting the ability to generate ROQ bases for emerging waveform models or extended parameter ranges.

In this work, we present a modified ROQ construction strategy that leverages multibanding techniques \citep{Vinciguerra:2017ngf} to accelerate the basis construction without compromising the accuracy of the final likelihood evaluations. Multibanding exploits the fact that gravitational wave signals from compact binary inspirals are not uniformly sampled in time across all frequencies, as lower frequencies contain proportionally more signal cycles than higher frequencies. By using frequency-dependent resolution during the basis search, with finer spacing at low frequencies and coarser spacing at high frequencies, we reduce the number of frequency samples by nearly an order of magnitude while preserving the essential features needed to construct accurate reduced bases. As we were completing the work, we became aware of a related idea \citep{Morisaki:2021ngj}, which applies multibanding to preexisting ROQ bases, to reduce ROQ basis file size and accelerate ROQ weight calculations and likelihood evaluations. In contrast, our approach implements multibanding during the ROQ basis construction itself, addressing the current computational bottleneck for long-duration signals by accelerating the greedy basis search.

We apply this multibanded construction method to generate ROQ bases for the \texttt{IMRPhenomXAS\_NRTidalV3} waveform model in the SSM regime, a parameter space of particular interest for searches targeting primordial black hole binaries and other exotic compact objects. Our results demonstrate that the multibanded approach reduces basis construction time by a factor of 5 to 20 times while simultaneously decreasing the basis size by 20-30{\%}. We validate these bases through comprehensive likelihood comparisons and full parameter estimation injection studies, confirming that the multibanded construction introduces no accuracy degradation in recovered source parameters.

The structure of this paper is as follows. In Sect. \ref{sec:2}, we review the fundamentals of gravitational wave parameter estimation, the ROQ formalism, multibanding techniques, and our multibanding implementation strategy in ROQ construction. Sect. \ref{sec:3} presents our construction results, and likelihood validation tests. In Sect. \ref{sec:4} we present parameter estimation comparisons between the PyROQ method and the modified PyROQ with multibanding techniques. We conclude in Sect. \ref{sec:5} with a discussion of the implications for future gravitational wave analyses and potential extensions of this methodology.


\section{Methodology} \label{sec:2}

\subsection{Gravitational Wave Inference}\label{2.1}

   Gravitational wave inference is performed to find the probability density function for a set of source parameters $\vec{\theta}$ used to model a gravitational wave signal depending on the detectors' observed strain data, $d$. This PDF or \textit{posterior} can be defined using Bayes Theorem:
   
   \begin{equation}
    \label{eq1}
      p(\theta|d) \propto \mathcal{L}(d|\vec{\theta})\pi(\vec{\theta}) \,,
   \end{equation}
   where $\mathcal{L}(d|\vec{\theta})$ is the \textit{likelihood} function of the data given the source parameters and $\pi(\vec{\theta})$ is the \textit{prior} probability for the source parameters. The highest computational cost is in evaluating the likelihood.

   The detector strain can be modelled as a combination of a gravitational wave signal, $h(\vec{\theta})$, and noise, $n$, such that $d = h(\vec{\theta})+n$ \citep{Likelihood}. Thus, we define the log-likelihood function as 

   \begin{equation}
   \label{eq2}
   \begin{aligned}
       \log \mathcal{L}(d|\vec{\theta})&=-\frac12(d-h(\vec{\theta}), d-h(\vec{\theta})) \\
       &=-\frac12(d,d)+(d, h(\vec{\theta}))-\frac12(h(\vec{\theta}),h(\vec{\theta})).
   \end{aligned}
   \end{equation}

   \noindent The overlap integral $(\cdot,\cdot)$ is defined as 

   \begin{equation}
   \label{eq3}
       (d, h(\vec{\theta}))=4\Re\Delta f\sum_{k=1}^L\frac{\tilde{d}^*(f_k)\tilde{h}(f_k;\vec{\theta})}{S_n(f_k)},
   \end{equation}

   \noindent where $S_n(f_k)$ is the power spectral density (PSD) of the detectors' noise.
   We can approximate the number of sampling points $L$ over an observation time $\tau=1/\Delta f$ as $L \approx \texttt{int}([f_{\texttt{high}}-f_{\texttt{low}}]\tau)$, where $f_{\texttt{low}}$ to $f_{\texttt{high}}$ is the detector frequency range. Longer duration signals increase the number of terms in Eq. \ref{eq3}, and more complex waveforms with a greater number of parameters $\vec{\theta}$ need more extensive sampling; more evaluations of Eq. \ref{eq3} are required. The computational costs of this impose a bottleneck on gravitational wave inference.

\subsection{Reduced Order Quadratures}

\begin{figure*}[h]
    \centering
    {\includegraphics[width=0.8\linewidth] {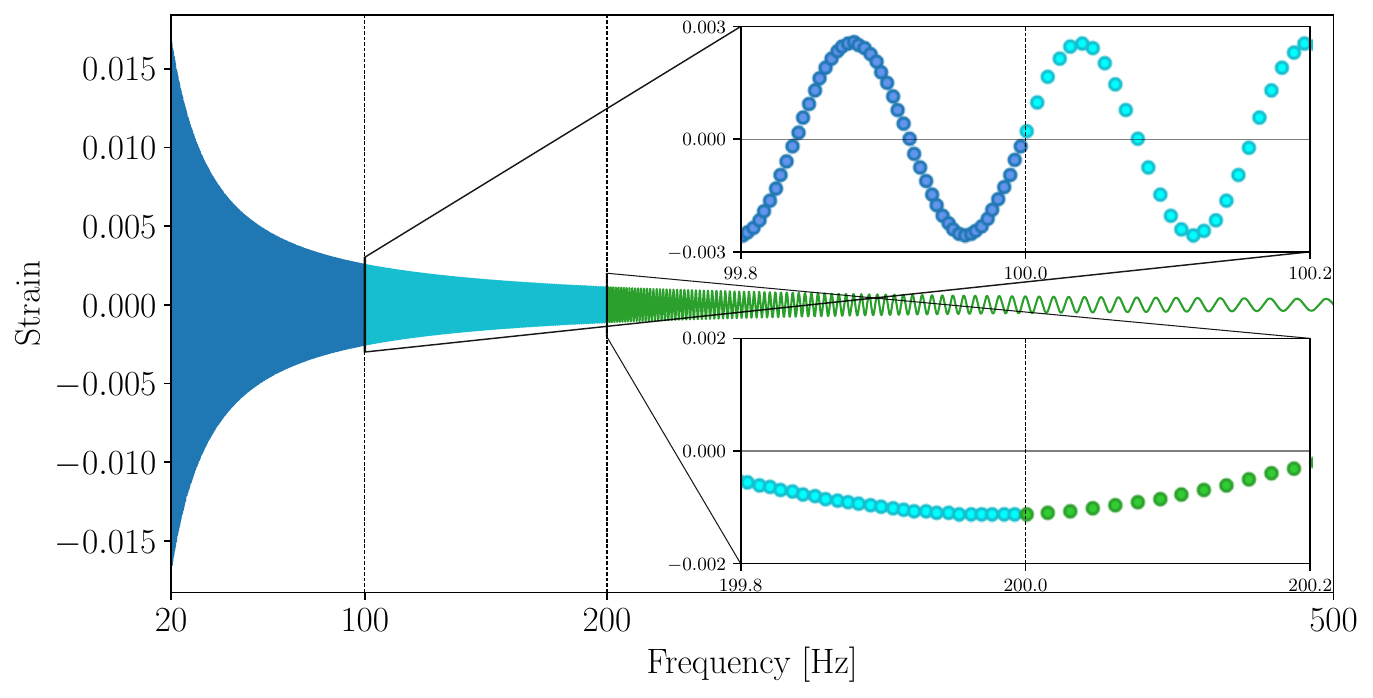}}
     \caption{Illustration of a normalized multibanded waveform in the frequency domain. The larger panel shows the full waveform, with the band boundaries marked using dashed vertical lines at $f = 100 ~\text{Hz}$, and $200~\text{Hz}$. The multiple colors represent the different $\Delta f$ used across each band. The upper and lower smaller panels show a zoomed-in view of the same waveform at the $f = 100~ \text{Hz}$ and $f = 200~ \text{Hz}$ band boundaries, respectively, highlighting the change in $\Delta f$ across bands. For both bands, the frequency spacing doubles.}
      \label{figMB}
\end{figure*}

In this section, a brief overview of the ROQ rule, along with its construction, is given. More detailed explanations can be found in \citep{Smith:2016qas}.
A gravitational waveform, $h(f_i;\vec{\theta})$, and its modulo squared, $|h(f_i;\vec{\theta})|^2$, can be represented by empirical interpolants, given by

\begin{subequations}
\begin{align}
\label{eq4a}
  &h(f_i;\vec{\theta}) \approx \sum_{j=1}^{N_\text{L}}B_j(f_i)h(F_j; \vec{\theta}), \\
  &|h(f_i;\vec{\theta})|^2 \approx \sum_{k=1}^{N_\text{Q}}C_k(f_k)|h(\mathcal{F}_k; \vec{\theta})|^2,
  \label{eq4b}
\end{align}
\end{subequations}

\noindent where $B_j(f_i)$, $C_k(f_i)$ are the reduced  (RBs), and $\{F_j\}_{j=1}^{N_{\text{L}}}$, $\{\mathcal{F}_k\}_{k=1}^{N_{\text{Q}}}$ are the interpolant nodes. By substituting Eq. \ref{eq4a} and \ref{eq4b} into the log likelihood, Eq. \ref{eq2}, the likelihood can be approximated as

\begin{equation}
    \log \mathcal{L} \approx -\frac12(d,d)+(d,h(\vec{\theta}))_{\text{ROQ}}-\frac12(h(\vec{\theta}),h(\vec{\theta}))_{\text{ROQ}},
\end{equation}

\noindent where the overlaps are calculated using precomputed quadrature weights.





Once found, the number of terms required to solve the ROQ likelihood is $N_{\text{L}} + N_{\text{Q}}$, leading to an $L/(N_{\text{L}} + N_{\text{Q}})$ speedup. The aim during ROQ construction is to minimize $N_{\text{L}} + N_{\text{Q}}$ whilst maintaining a desired threshold accuracy.

A brief overview of how the ROQ basis elements are found using the PyROQ code will be provided, with more detailed explanations available in \citep{PyROQ}. The process begins with the generation of a large training set of waveform parameters $\vec{\theta}$. After an initial basis element is found, an iterative process occurs where the parameters associated with the waveform with the highest maximum empirical interpolation error are added to the basis. This process terminates once the entire training set falls below a threshold empirical interpolation error. The dominant computational cost occurs at this stage, as it requires a high volume of generated waveforms and inner product evaluations. This means that for longer or more complex waveforms, computational cost can be high.

\subsection{Multibanding}\label{sec:2.3}

In traditional PE methods, for a signal of duration $\tau$, the frequency resolution is fixed at $\Delta f = \tau^{-1}$, with the total number of samples given by $N = (f_{\text{max}}-f_{\min})\tau$. The amount of time, $t$, an inspiral signal spends at frequency, $f$, is given by \citep{Chirping}

\begin{equation}
\label{eq5}
    t(f) = 5(8\pi f)^{-8/3}\mathcal{M}_c^{-5/3},
\end{equation}

\noindent where $\mathcal{M}_c$ is the chirp mass of the source object. From this equation, we find that a majority of the waveforms cycles occur at lower frequencies. Therefore, using a uniform frequency grid will result in certain parts of the signal being oversampled. Multibanding takes advantage of this by having the frequency resolution be dependent on the frequency itself. More specifically, the full frequency range is split into multiple bands, each of which has different but uniform frequency resolutions. A visual representation of this can be found in Figure \ref{figMB}. For this work, we construct multibands using the following structure, built on the base frequency spacing $\Delta f_0 = 1/\tau$,

\begin{equation} \label{eq:mb}
\Delta f(f) =
\begin{cases}
\Delta f_0, & f_{\min} \le f < f_1 \\
2\,\Delta f_0, & f_1 \le f < f_2 \\
\multicolumn{1}{c}{\vdots} & \multicolumn{1}{c}{\vdots} \\
2^{N}\,\Delta f_0, & f_{N} \le f \le f_{\max}.
\end{cases}
\end{equation}

\noindent The frequency spacing increases by a scaling factor of 2 between successive bands. We experimented with greater scaling factors, which led to notable decreases in the likelihood error. We thus found the $2^n$ scale to be the best balance between waveform reconstruction accuracy and construction efficiency.

\begin{table*}[h!]
\caption{Comparison in ROQ construction using the standard and multibanded construction.}
\centering
\begin{tabular}{llcccccccccc}
\hline \hline
\noalign{\smallskip}
\multirow{2}{*}{Bases} & \multirow{2}{*}{Multibanding} & \multirow{2}{*}{\# Frequency Samples} & \multicolumn{2}{c}{Basis Size} & \multicolumn{2}{c}{Construction Time [h]}  \\
& & & Linear & Quadratic & Linear & Quadratic \\
\hline
\noalign{\smallskip}
256s & No & \num{916481} & 303 & 98 & 105.78 & 37.35  \\
256s & Yes & \num{110081} & 233 & 68 & 7.99 & 5.85  \\
\noalign{\smallskip}
\hline
\bottomrule
\end{tabular}
\label{tab:con_label}
\end{table*}

\subsection{Implementing Multibanding in ROQ Construction}

For this work, we implement multibanded waveforms solely during the ROQ greedy basis search stage. The final likelihood evaluations remain unchanged compared to the standard ROQ method. The modified method is given here:

\begin{enumerate}
  \item The first stage generates the multibanded frequency array. A key detail is that the same multibanded array must be used for all waveforms across the desired parameter range for which the ROQs will be built. The chosen multibands must be flexible enough to cover this space, whilst still minimizing the number of sampling points. As mentioned in Sect. \ref{sec:2.3}, bands in the from of Eq. \ref{eq:mb} were found to meet both these requirements.
  \item The next step is to generate the training dataset as usual. Each data point consists of a set of parameter values, $\vec{\theta}_i$, which represent a waveform found within the desired parameter range.
  \item The PyROQ reduced basis search algorithm, as found in \citep{PyROQ}, is performed. The key difference is that the training waveforms are evaluated using the multibanded frequency array. The parameter values of the RB elements, $\vec{\theta}_i^{\text{MB}}$, are stored.
  \item Using the multibanded parameter values $\vec{\theta}_i^{\text{MB}}$, the RB is reconstructed using the full frequency array, and standard PyROQ construction continues until completion.
\end{enumerate}

\noindent Since the modification affects only the greedy basis search stage, the final MB constructed ROQ structure is identical to that produced by the standard PyROQ , with the final bases using the full frequency array. This means these ROQs remain fully compatible with current PE software.

\section{ ROQ and Multibanded ROQ Comparison}\label{sec:3}

\subsection{Basis Construction}\label{sec:3.1}

In this section, we compare the performances of our multibanded construction method with the standard construction method found in PyROQ. We then show how our method provides both ROQ construction speed-up and basis size reduction with negligible loss in accuracy.

For this, we use the \texttt{IMRPhenomXAS\_NRTidalv3} waveform model \citep{XASwaveform}, as implemented in the LIGO Algorithm Library (LAL) \citep{lalsuite}. This phenomenological model approximates inspiral-merger-ringdown signals for aligned spin systems. It is also capable of incorporating tidal effects, such as those found in BNS and NSBH systems. This waveform was chosen as it produces accurate long-duration signals in SSM ranges, which is an area that is proving computationally intensive for current ROQ construction methods.

PyROQ was used to construct both bases, with identical parameter ranges used in both. For the 256~s signal, the chirp mass range was $0.995\ M_\odot$ to $1.005\ M_\odot$, with mass ratios ranging from 1 to 4.1. The dimensionless spin magnitude, $a_{1,2}$, ranging from 0 to 1, and the tidal deformabilities, $\Lambda_{1,2}$, ranging from 0 to 5000. The frequency range is 20 Hz to 3600 Hz. The dataset for the standard ROQ used a uniform frequency step size of $\Delta f_0=1/256\ \text{Hz}$, whereas for the multibanded method, the step size used is given by

\begin{equation}
\Delta f(f) =
\begin{cases}
\Delta f_0, & 20\ \text{Hz} \le f < 100\ \text{Hz} \\
2\,\Delta f_0, & 100\ \text{Hz} \le f < 200\ \text{Hz} \\
4\,\Delta f_0, & 200\ \text{Hz} \le f < 500\ \text{Hz} \\
8\,\Delta f_0, & 500\ \text{Hz} \le f < 1000\ \text{Hz} \\
16\,\Delta f_0, & 1000\ \text{Hz} \le f \le 3600 \ \text{Hz}.
\end{cases}
\end{equation}

\noindent This means each multiband waveform used during construction has \num{110081} frequency samples, compared to the \num{916481} used in the standard method, a $\sim 9$ times decrease. The threshold tolerance is $10^{-5}$ and $10^{-10}$ for the linear and quadratic bases respectively. A training dataset of 1 million waveforms was used in both constructions. Both bases were constructed using 32 CPUs, with the linear bases using 256GB RAM and the quadratic bases using 64GB RAM on the LIGO data grid.

\begin{figure}[ht!]
    \centering
    {\includegraphics[width=1.0\linewidth]{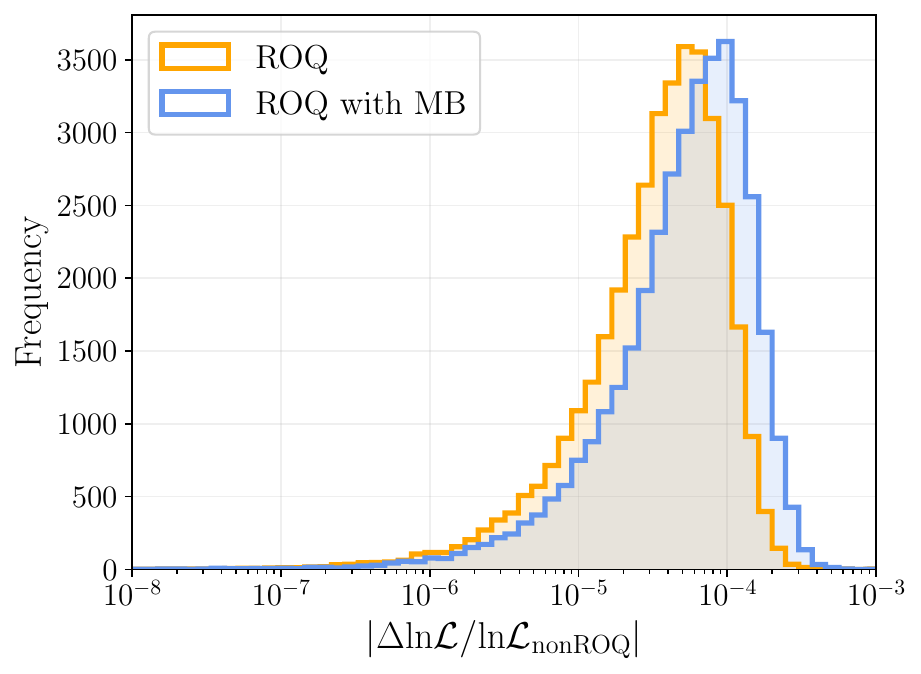}}
     \caption{The distributions of the likelihood errors between the full likelihood and the ROQ likelihood (orange) and the multiband ROQ likelihood (blue), respectively, for \num{38000} randomly drawn injected waveforms. The samples were drawn from the parameter space used to construct the 256~s bases, as described in Section 2.3.  A majority of samples were found with likelihood errors less than $6\times 10^{-4}$. }
      \label{figLE}
\end{figure}

\begin{figure*}[ht!]
    \centering
    {\includegraphics[width=0.8\linewidth]{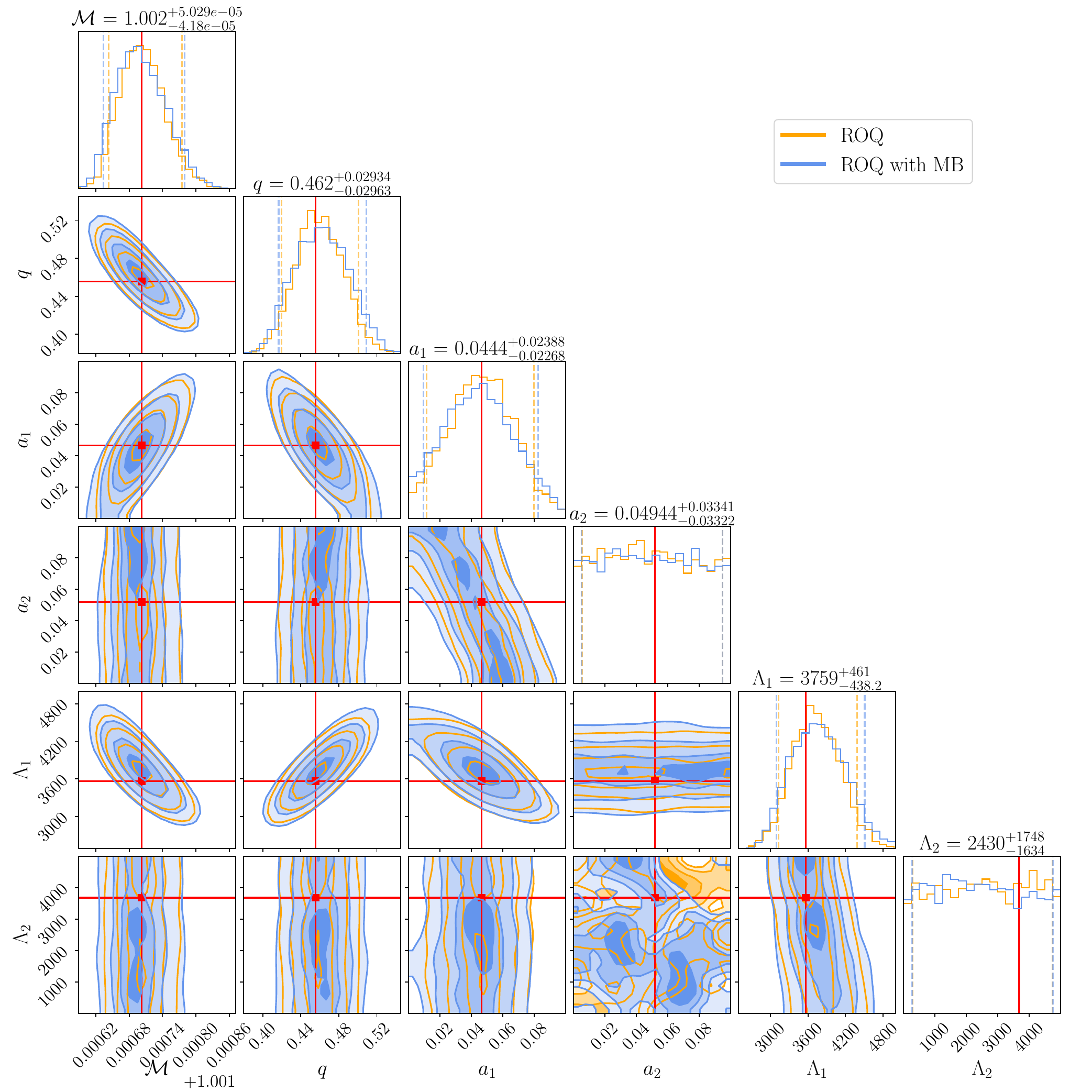}}
     \caption{Corner plots comparing the posterior distributions between the standard PyROQ method (orange) and the multibanded construction method (blue). The vertical red lines show the injected parameter values. The dashed lines represent the 5\% and 95\% credible intervals.}
      \label{figCor}
\end{figure*}

The results for the two constructions can be found in Tab. \ref{tab:con_label}. Comparing the 256~s bases, we see an approximate 30\% decrease in the basis size (303 versus 233 for linear and 98 versus 68 for quadratic). The biggest change is found in the overall construction time with the linear basis constructed in 7.99~h rather than 105.78~h, a $\sim$ 13 times speed up, and the quadratic constructed in 5.83~h rather than 37.35~h, a $\sim$ 6 times speed up. More exact values are given in Tab. \ref{tab:Speedup}.

These results show that introducing multibanding during the ROQ basis search both speeds up the construction whilst decreasing the overall basis size. For long-duration signals, such as the 256~s ones used here, this improvement is especially apparent with the construction time decreasing from a few days to just several hours, enabling the construction of these previously impractical ROQs.

\begin{table}[ht!]
\centering
\begin{tabular}{ccc}
\toprule
\toprule
Basis Type & Basis Size Change & Construction Speedup \\
\midrule
Linear & $-23.1\%$ & $13.2\times$ \\
Quadratic & $-30.6\%$ & $6.38\times$ \\

\bottomrule
\bottomrule
\end{tabular} \caption{\label{tab:Speedup}Table showing the basis size decrease and basis construction time speed up, going from the standard ROQ construction to the multibanded ROQ construction. This is for the bases constructed in Sect. \ref{sec:3.1}.}
\end{table}

\subsection{Likelihood Error}\label{Sec:likeCom}

We now calculate the likelihood error of our multiband constructed \texttt{IMRPhenomXAS\_NRTidalv3} ROQ relative to the full likelihood and the basis built using the standard PyROQ method. This is done using BILBY \citep{Ashton2019}.

To do this, we generated \num{38000} samples from our parameter space defined in Sect. \ref{Sec:likeCom}. The percentage error between the difference in likelihoods and the full likelihood is calculated. The distribution for the likelihood error between the multibanded constructed ROQs and the full likelihood is shown in Fig. \ref{figLE}. From this, we see that the majority of the likelihood errors do not exceed $6\times 10^{-6}$, with only 43 outliers falling above $10^{-3}$. Comparing the likelihood error with the standard ROQ constructed bases, we see a slight increase in the overall error. However, we will see later that this error increase proves negligible during a PE run.

\begin{table*}[ht!]
\centering
\caption{\label{tab:injection_and_pe_priors} The parameter values for the injected waveform shown in Fig. \ref{figCor}, and the distribution and ranges used as the prior for the PE injection analysis.}
\label{tb:distributions}
\begin{tabular}{l*{6}{l}l}
\hline\hline
Parameter (Symbol) [Unit] \quad  & Injection values \quad  & PE prior
\\ \hline
Source-frame chirp mass  ($\mathcal{M}_{\rm c})\ [M_{\odot}]$ & 
$1.002$ & Uniform [0.995, 1.005]  \\
Source-frame mass ratio ($q$) &  
0.4556 & Uniform [0.244, 1] \\
Dimensionless primary NS spin  ($a_{1}$)  & 0.04654 & Uniform [0, 0.1] \\
Dimensionless secondary NS spin ($a_{2}$)  & 0.05186 & Uniform [0, 0.1] \\
Luminosity distance  ($d_L)\ [{\rm Mpc}]$ & 24.27  & \text{Square power law} [10, 50] \\
Right ascension  ($\alpha$) [radian] & 0.9453 & Uniform [0, $2\pi$] \\
Declination ($\delta$) [radian] & 5.113 & Uniform Cosine \\
Inclination angle ($\theta_{\text{JN}}$) [radian] & 0.5854 & Uniform Sine [0, $\pi$]\\
Polarization  ($\Psi$) [radian] & 0.4481 & Uniform [0, $\pi$] \\
Tidal deformability of primary NS ($\Lambda_1$) & 3566 & Uniform [0, 5000] \\
Tidal deformability of secondary NS ($\Lambda_2$) & 3683 & Uniform [0, 5000] \\
\hline\hline
\end{tabular}
\end{table*}

\section{Parameter Estimation Validation}\label{sec:4}

\subsection{Gravitational Wave Injections}

To validate the accuracy of our multiband constructed basis, we performed a set of twenty injection parameter estimation runs. The signals were simulated based on the prior values given in Tab. \ref{Sec:likeCom}, and injected into Gaussian noise, using A+ projected PSDs. The source parameters were recovered using both the standard PyROQ basis  and the basis constructed with the multiband construction method. The parameter estimation was performed with BILBY, using the \texttt{dynesty} sampler with \texttt{nlive}=1000 and \texttt{nact}=10. Twenty injection runs were performed, with all jobs running on the LIGO data grid using 16 CPUs and 16~GB of RAM each.

\subsection{A PE Comparison Example}

We showcase one example injection signal, with chirp mass $1.002~M_{\odot}$, mass ratio 0.4556, dimensionless spins $a_{1,2}=\{0.04654, 0.05186\}$, and EOS agnostic tidal deformabilities $\Lambda_{1,2}=\{3566, 3683\}$. Both the injected values and the parameter estimation priors for the injected signal are given in Tab. \ref{tab:injection_and_pe_priors}. For the injection in Fig. \ref{figCor}, the PE run time using the multiband constructed ROQs was 2.99~h compared to 2.93~h using the standard constructed ROQs.

The corner plot shows the results for the intrinsic source parameters, namely the chirp mass $\mathcal{M}$, mass ratio $q$, spin magnitudes $a_{1,2}$, and tidal deformabilities $\Lambda_{1,2}$. Across all parameters, we see clear agreement with both methods. No obvious biases are observed, and the credible intervals are consistent through both methods. 

\subsection{Population Study}\label{sec:4.3}

In Tab. \ref{table:4}, we find the \textit{root mean square error} (RMSE) between the median posterior and injection values with all 20 injected runs. For the MB constructed ROQs, we find comparable performance across all parameters, with modest improvements for phase-dominant parameters such as the chirp mass, $\mathcal{M}_c$, and mass ratio $q$. Conversely, we find a slight increase in error for tidal parameters, which contribute strongest during the higher-frequency, late merger-stage \citep{tidalEffects}. This is consistent with the use of multi-banded waveforms during greedy basis search, which biases basis construction towards regions in the waveform which are more parameter dependent.

\begin{figure}[ht!]
    \centering
    {\includegraphics[width=1.0\linewidth]{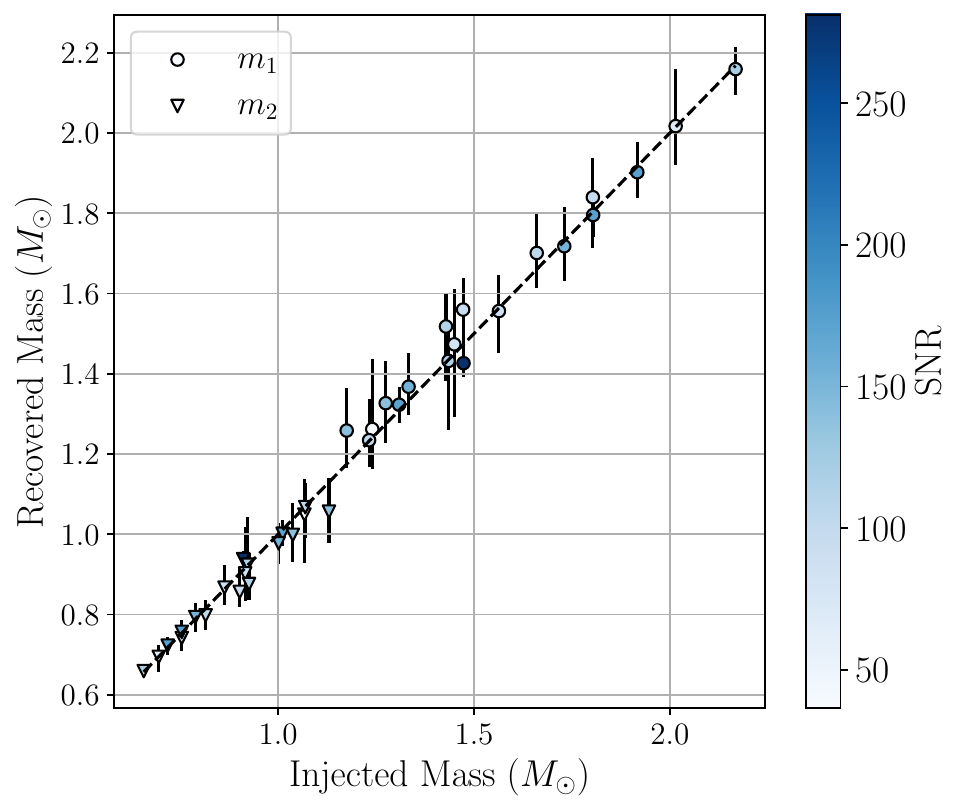}}
     \caption{Comparison of injected versus recovered component masses, for the twenty injections given in Sec. \ref{sec:4}. The x-axis shows the injected mass values, and the y-axis shows the recovered median posterior mass value. The error bars show the $90\%$ credible intervals. The dashed line shows exact parameter recovery.}
      \label{figMassPop}
\end{figure}

Fig. \ref{figMassPop} shows the recovered individual mass components $m_{1,2}$, where $m_1 >m_2$, derived from the recovered chirp masses and mass ratios for the twenty injection runs, using the MB constructed ROQs. For all 40 masses, the injected mass value fell within the $90\%$ credible interval of the recovered mass.





\begin{table}[ht!]
\centering
\begin{tabular}{lll}
\toprule
\toprule
\multirow{2}{*}{Parameter} & \multicolumn{2}{c}{RMSE}  \\

 & ROQ & ROQ with MB \\
\midrule
$\mathcal{M}_c [M_{\odot}]$ & $2.560\times10^{-5}$ & $2.436\times10^{-5}$  \\
$q$ & $4.293\times10^{-2}$ & $4.163\times10^{-2}$ \\
$a_1$  & $1.272\times10^{-2}$ & $1.306\times10^{-2}$ \\
$a_2$  & $1.428\times10^{-2}$ & $1.406\times10^{-2}$ \\
$d_L [\text{Mpc}]$  & $ 3.664$ & $3.629$ \\
$\alpha [\text{radian}]$  & $9.670\times10^{-3}$ & $9.664\times10^{-3}$ \\
$\delta [\text{radian}]$  & $4.836\times10^{-3}$ & $4.834\times10^{-3}$ \\
$\theta_{\text{JN}} [\text{radian}]$  & $0.1685$ & $0.1679$ \\
$\Psi [\text{radian}]$  & $0.9700$ & $0.7582$ \\
$\Lambda_1$  & $191.9$ & $219.9$ \\
$\Lambda_2$  & $944.3$ & $978.2$ \\

\bottomrule
\bottomrule
\end{tabular} \caption{\label{tab:InjectRecov}Table showing the root mean square error between the median posterior values and the injected values of the 20 injections in Sect. \ref{sec:4.3}, comparing the standard constrcuted ROQs (denoted as ROQ) to the multibanded constructed ROQs (denoted as ROQ with MB). $\Delta$RMSE is the percentage RMSE difference when using the multibanded constructed ROQs over the standard constructed ROQs.}\label{table:4}
\end{table}

\section{Conclusions}\label{sec:5}
We have described a modification to the algorithm implemented in PyROQ, which significantly speeds up the construction of reduced bases for gravitational wave waveforms, by utilizing multibanded waveforms during the basis search.

We used our multiband ROQ construction method to construct an SSM basis for the \texttt{IMRPhenomXAS\_NRTidalv3} waveform model. Comparing this to a ROQ basis constructed using the standard PyROQ method over identical parameter ranges, we found a $\sim 30\%$ decrease in basis size, and a $\sim 13$ and $\sim 9$ times decrease in construction time for the linear and quadratic bases, respectively. We also demonstrated our basis in a PE run, and found that the altered method did not affect the final PE results relative to the standard PyROQ constructed bases.

The constructed bases for \texttt{IMRPhenomXAS\_NRTidalv3} are the first to be built in SSM ranges, and demonstrate the capabilities of this new method. This will significantly reduce the computational cost for any parameter estimation of these longer-duration signals. Future work will focus on further implementing this method, particularly in constructing ROQs for long-duration waveforms for the next-generation detectors. Further testing will also be performed to test the methods capability for precessing and higher mode waveforms.

\begin{acknowledgements}
We are grateful for the computational resources provided by LIGO Laboratory and the Leonard E Parker Center for Gravitation, Cosmology and Astrophysics at the University of Wisconsin-Milwaukee and supported by National Science Foundation Grants PHY-0757058, PHY-0823459, PHY-1626190, and PHY-1700765. We especially thank the computing resources provided by Digital Research Alliance of Canada through two consecutive grants, the DRAC RPP $\#$1012 and the Compute Canada Allocation Award $\#$696 to the University of British Columbia. We also thank the Hawk supercomputing system provided by Cardiff University. This material is based upon work supported by NSF's LIGO Laboratory which is a major facility fully funded by the National Science Foundation. 
\end{acknowledgements}

%

\bibliographystyle{aa}
\bibliography{bibliography.bib}

\FloatBarrier 

\end{document}